\documentclass[apj]{emulateapj}
\lefthead{TSUJIMOTO \& BEKKI}
\righthead{TWO-COMPONENT GALACTIC BULGE}

\def\ltsima{$\; \buildrel < \over \sim\;$}
\def\ltsim{\lower.5ex\hbox{\ltsima}}
\def\gtsima{$\; \buildrel > \over\sim \;$}
\def\gtsim{\lower.5ex\hbox{\gtsima}}
\def\ms{$M_{\odot}$ }
\def\msp{$M_{\odot}$}

\slugcomment{Accepted for publication in ApJ}

\begin{document}
\title{Two-Component Galactic Bulge Probed with Renewed Galactic Chemical Evolution Model}

\author{Takuji Tsujimoto$^{1}$ and Kenji Bekki$^{2}$}

\affil{$^1$National Astronomical Observatory of Japan, Mitaka-shi,
Tokyo 181-8588, Japan; taku.tsujimoto@nao.ac.jp \\
$^2$ICRAR, M468, The University of Western Australia 35 Stirling Highway, Crawley Western Australia, 6009
}

\begin{abstract}
Results of recent observations of the Galactic bulge demand that we discard a simple picture of its formation, suggesting the presence of two stellar populations represented by two peaks of stellar metallicity distribution (MDF) in the bulge. To assess this issue, we construct Galactic chemical evolution  models that have been  updated in two respects: First, the delay time distribution (DTD) of type Ia supernovae (SNe Ia) recently revealed by extensive SN Ia surveys is incorporated into the models. Second, the nucleosynthesis clock, the $s$-processing in asymptotic giant branch (AGB) stars, is carefully considered in this study. This novel model first shows that the Galaxy feature tagged by the key elements, Mg, Fe, Ba for the bulge as well as thin and thick disks is compatible with a short-delay SN Ia. We present a successful modeling of a two-component bulge including the MDF and the evolutions of [Mg/Fe] and [Ba/Mg], and reveal its origin as follows. A metal-poor component ($<$[Fe/H]$>$$\sim$-0.5) is formed with a relatively short timescale of $\sim$1 Gyr. These properties are identical to the thick disk's characteristics in the solar vicinity. Subsequently from its remaining gas mixed with a gas flow from the disk outside the bulge, a metal-rich component  ($<$[Fe/H]$>$$\sim$+0.3) is formed with a longer timescale ($\sim$ 4 Gyr) together with a top-heavy initial mass function that might be identified with the thin disk component within the bulge. 
\end{abstract}

\keywords{Galaxy: bulge --- Galaxy: abundances --- Galaxy: evolution --- stars: abundances}

\section{Introduction}

The Galactic bulge is classified as a boxy bulge that is associated with bars and is likely generated through disk instability processes \citep[e.g.,][]{Kuijken_95, Bureau_99}. Its kinematics and surface brightness profile are indeed shown to be cylindrical rotation \citep{Howard_09, Shen_10} and be near-exponential rather than $r^{1/4}$ \citep{Kent_91}, respectively as evidence for boxy bulge. This view suggests that the formation of the Galactic bulge is closely connected to disk evolution, resulting in a longer timescale for its formation than that expected for classical bulges, considered to be merger products as a result of hierarchical galaxy formation in the cold dark matter (CDM) Universe \citep{Aguerri_01, Scannapieco_03}. Thus, this views favors a recent finding of a large age span of $\sim$2-13 Gyr among microlensed turn-off bulge stars \citep{Bensby_11}, though the color-magnitude diagrams of the Galactic bulge show no clear signature of the presence of a young stellar population \citep[e.g.,][]{Zoccali_03, Clarkson_08}. 

In general, it is expected that a disk instability forming the bulge induces a vertical mixing, which leads to erasing a metallicity gradient along a minor axis. This contradicts the observed result showing a clear [Fe/H] gradient \citep{Zoccali_08}. This problematic issue is solved by introducing two components, i.e., thick and thin disks, as the origin of Galactic bulge \citep{Bekki_11a}. Their scenario is as follows: The first disk is disturbed by an ancient minor merger, which induces a vertical growth of the disk and transforms it into a thick disk, and subsequently the thin disk starts to form with an accompanying bar formation in the central region. In such a two-component disk, a vertical mixing induced by a bar buckling functions incompletely in a sense that the high latitude region in the thick disk is not well mixed. In the end, the resultant bulge shows a vertical metallicity gradient as well as a cylindrical rotation, both of which are compatible with the observed results.

Two components of the Galactic bulge were first proposed by \citet{Babusiaux_10} in terms of chemistry and kinematics. They find the metal-poor population kinematically associated with an old spheroidal or a thick disk and the metal-rich population with a bar-like kinematics. Furthermore, very recently, a double-peaked metallicity distribution (MDF) of bulge stars has been reported with two different tracers, i.e., red clump giants \citep{Hill_11} and microlensed dwarf stars \citep{Bensby_11}. Two results basically point to the same conclusion that the metallicity of one peak is metal-poor at [Fe/H]$\sim$ -0.6 - -0.3 and the other metal-rich at [Fe/H]$\sim$+0.3.

In view of chemical evolution, the MDF should be examined together with the time evolution of the elemental abundance pattern generally described as [X/Fe] vs.~[Fe/H].  One aspect of the converged knowledge on this for the bulge is that the [$\alpha$/Fe] ratio decreases with an increasing [Fe/H] for  the metal-rich regime \citep[e.g.,][]{Alves_10, Hill_11,Bensby_11}, which bears witness to the presence of contribution from type Ia supernovae (SNe Ia) to chemical enrichment in the bulge. Here we should highlight recent results regarding  the delay time distribution (DTD) of SNe Ia yielded by the studies on the SN Ia rate in distant and nearby galaxies \citep{Mannucci_06, Sullivan_06, Totani_08, Maoz_11}. Their findings dramatically shorten the SN Ia's delay time, compared with its  conventional timescale of $\sim$1 Gyr, which will have a significant impact on Galactic chemical evolution. 

This renewed picture of a SNIa clocking should be tied up with another nucleosynthesis clock, the $s$-process operating in an asymptotic giant branch (AGB) star. These two different delayed-timings to release the nucleosynthesis product will create variation in stellar abundances among the elements such as Mg, Fe, and Ba, from which we can decipher  the evolutionary pathway to the present through detailed modeling. In this paper, incorporating the new SN Ia DTD and AGB yield into the Galactic chemical evolution (GCE) model, we explore the chemical evolution of the Galactic bulge, highlighting its two components. To strengthen the rigidness of the model, we first complete the reproduction of the chemical feature of the Galactic thin and thick disks.

\section{Two Nucleosynthesis Clocks}

Chemical enrichment at an early stage is promoted by the product from short-lived massive stars through type II SNe (SNe II), and subsequently with a time delay, heavy elements from SNe Ia and AGB stars come to an entry. As a result, the relative contribution from SNe Ia or AGB stars to the ISM in comparison with SNe II, which is imprinted in individual stellar abundances of long-lived stars, can be utilized as an age-dating for the stars. The updates of the input for these delayed contributors are briefly described.

\subsection{SNe Ia}

It is difficult to estimate the evolutionary timescale of SN Ia progenitors from purely theoretical arguments due to difficulties in identifying the binary companions of SN Ia's progenitors. Therefore, the break in the [$\alpha$/Fe] ratio seen in the solar neighborhood stars due to the switchover of the major Fe source from SNe II to SNe Ia at [Fe/H] $\sim$-1 had long been the only information available for deducing the delay time of SNe Ia by theoretically estimating the elapsed time until [Fe/H]$\sim$-1. This assessment indeed leads to a considerably long delay time of $\sim$1-1.5 Gyr \citep[e.g.,][]{Pagel_95, Yoshii_96}. However, recently we have entered a new phase in our understanding of SN Ia delay time. Its distribution, i.e.,  the DTD, is revealed by the survey of SN Ia rate in extragalaxies, the individual galaxy ages of which are deduced. Surprisingly, it turns out that young progenitors for SNe Ia are dominant \citep{Mannucci_06, Sullivan_06}, and that the DTD is proportional to $t_{\rm delay}^{-1}$ for the approximate range of 0.1 Gyr $\leq t_{\rm delay} \leq$10 Gyr \citep{Totani_08, Maoz_10}. This implies that about 70\% of SNe Ia explodes with a time delay within 1 Gyr. Such a form of DTD can be predicted by the theoretical models for the progenitor of SNe Ia: a double-degenerate (DD) scenario, i.e., merging of double C+O white dwarfs with a combined mass surpassing the Chandrasekhar mass limit \citep[e.g.,][]{Yungelson_00, Greggio_05} or a single-degenerate (SD) scenario, i.e., accretion of hydrogen-rich matter via mass transfer from a binary companion \citep{Hachisu_08}. In this study, we apply this new SN Ia DTD to Galactic chemical evolution, and for its normalization we assume the mass of the primary component of binary, which eventually produce SNe Ia through accretion or merging, in the range of $3-8$ \msp. In the end, the SN Ia rate is calculated so that  the fraction of the stars that end up with SNe Ia for this mass range is 0.08, which is distributed with a DTD $\propto t_{\rm delay}^{-1}$ for $0.1\leq t_{ \rm delay}\leq10$ Gyr. Its fraction has been obtained through previous works as well as recheck by this study. In \S 3, we will see a fatal problem that we have noticed in a previous approach to try to determine the SN Ia DTD within the Galaxy thanks to advancement of the observed data.

\begin{figure}[t]
\vspace{0.2cm}
\begin{center}
\includegraphics[width=7cm,clip=true]{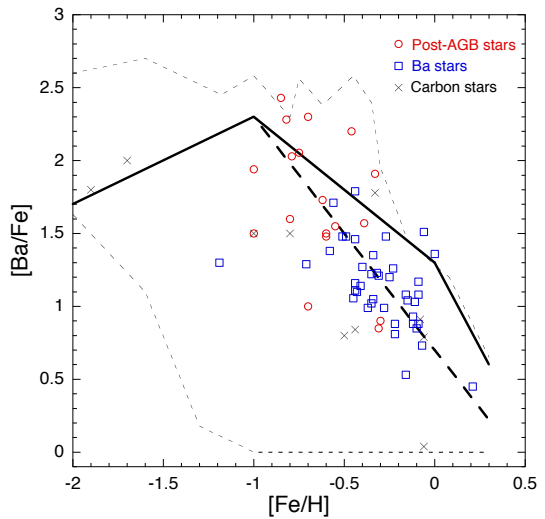}
\end{center}
\vspace{0.3cm}
\caption{Comparison of the theoretical [Ba/Fe] ratio expected for AGB stars as a function of [Fe/H] with the observed abundances of Post-AGB stars (red circles), Ba stars (blue squares), and Carbon stars (crosses). Solid lines denote the ratio of Ba yield synthesized in AGB stars to stellar metallicity used for model calculations. The Ba yield corresponding to dashed line showing a better fit to the observation is also applied to calculations (see Fig.3). Two dotted curves represent the maximal and minimal predictions from AGB models with a range of $^{13}$C-pocket efficiencies \citep{Busso_01}.
}
\end{figure}

\subsection{$s$-processing}

Low-mass AGB stars ($M <$3\msp) release the $s$-process elements during the thermally pulsing AGB phase \citep[e.g.,][]{Gallino_98}. Mainly due to large uncertainties in convective mixing and $^{13}$C-pocket efficiencies, the $s$-process nucleosynthesis allows a wide range for the level of a possible production. On the other hand, abundances of the surface of AGB stars can be directly compared with the nucleosynthesis results. Here we focus on the element, Ba. Figure 1 shows the observed abundances of Post-AGB stars, Ba stars, and Carbon stars \citep[][references therein]{Kappeler_11}. Note that all observed data reside in the theoretically allowable range  \citep{Busso_01} denoted by dashed lines. Here we determine the best empirical Ba yield as a function of stellar [Fe/H] so as to reproduce the Ba evolution of thin disk stars within an observationally acceptable range in terms of AGB abundances. For [Fe/H]$>$-1, the Ba yield given by dashed line which is fully consistent with the data fails to reproduce the chemical evolution of the thin disk. Alternatively, we adopt the Ba yield for stars with a mass of 1.5-3 \ms corresponding to the solid line which is rather close to the upper envelope of the observed [Ba/Fe]-[Fe/H] plane.

\begin{figure}[t]
\vspace{0.2cm}
\begin{center}
\includegraphics[width=7cm,clip=true]{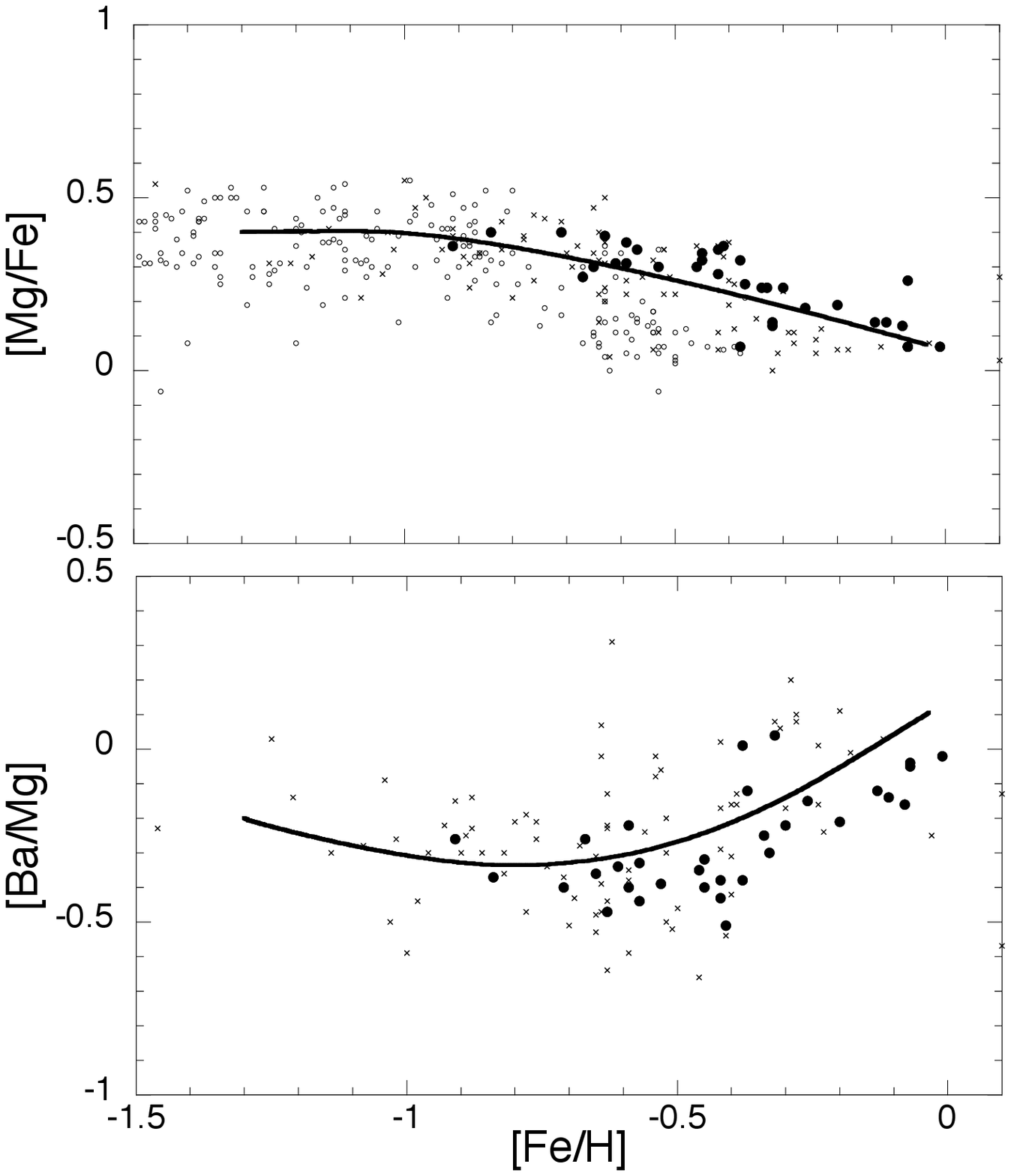}
\end{center}
\vspace{0.3cm}
\caption{Predicted features of chemical evolution in the thick disk, compared with the observed quantities.  {\it Top panel}: The [Mg/Fe]-[Fe/H] diagram. The observed data are taken from \citet{Bensby_05}, \citet{Ruchti_11},  and \citet{Venn_04}, denoted by filled circles, small circles, and small crosses, respectively. {\it Middle panel}: The [Ba/Mg]-[Fe/H] diagram. The [Ba/Mg] ratios from \citet{Venn_04} are shifted by +0.15 dex to broadly adjust their average to the level of data from \citet{Bensby_05}. 
}
\end{figure}

\section{Chemical Evolution of Disks}
Solar neighborhood stars are a mixture of stars belonging to different Galactic components, i.e., halo, thick disk, and thin disk. A growing understanding of the kinematic properties of individual stars thanks to Hipparcos data has enabled us to precisely select thick and thin disk stars \citep[e.g.,][]{Feltzing_03}. It then reveals that the observed break in [$\alpha$/Fe] is seen among thick disk stars (Fig.~2), and that all thin disk stars follow a decreasing [$\alpha$/Fe] trend (Fig.~3). Accordingly, our previous view that some fraction of thin disk stars populates a plateau of [$\alpha$/Fe] with respect to [Fe/H] has to be discarded. In this study, we examine the evolution of [Mg/Fe] and [Ba/Mg] (not [Ba/Fe] so as to prevent the effect of $s$-processing from being hidden by SN Ia contamination). In addition, we focus on the chemical evolution for [Fe/H]\ltsim 0 since the origin of metal-rich disk stars should be assessed with an extra evolution factor such as either stellar migration \citep{Roskar_08, Bekki_11b} or bulge winds \citep{Tsujimoto_07, Tsujimoto_10}.

We should note that Ba is synthesized through not only $s$-process but also $r$-process. The theoretical interpretation of abundance data on very metal-poor stars implies that the mass range for the $r$-process is 8-10 \ms \citep{Mathews_92, Ishimaru_99} as identified with the collapsing O-Ne-Mg core \citep{Wheeler_98}, 20-25 \ms \citep{Tsujimoto_00},  or 12-30 \ms \citep{Cescutti_06} with heavy weighing on lower mass progenitors. Here, as the site for $r$-process, we adopt the mass range of 20-25 \ms and the Ba yield deduced by \citet{Tsujimoto_00}. We do not have to be exact in the mass range because its variance influences only the behavior of $r$-process enrichment at a very early epoch, outside the focus of this study.

\begin{figure}[t]
\vspace{0.2cm}
\begin{center}
\includegraphics[width=7cm,clip=true]{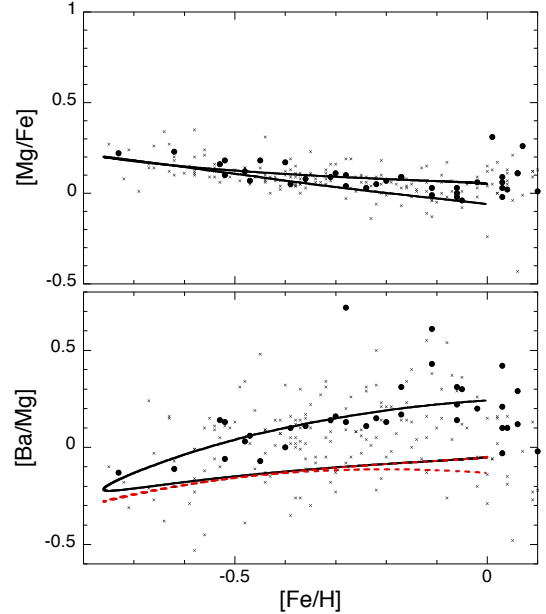}
\end{center}
\vspace{0.3cm}
\caption{Predicted features of chemical evolution in the thin disk. In the models, chemical evolution is assumed to start from the remaining gas after the formation of the thick disk, and the results are shown by solid curves. The dashed curve is the result calculated with the Ba yield, the metallicity dependence of which corresponds to the dashed line in Fig.1. The [Ba/Mg] ratios from \citet{Venn_04} are shifted by +0.15 dex as in Fig.~2. 
}
\end{figure}

Since the present GCE models are essentially the same as those adopted in our previous paper \citep[e.g.,][]{Tsujimoto_10, Bekki_11b}, we briefly describe the models. The basic picture is that the Galactic thin and thick disks were formed by the gas infall from outside the disk region. For the infall rate, we apply an exponential form with a timescale $\tau_{\rm in}$. Taking into account the relatively rapid formation of the thick disk and the presence of a G-dwarf problem for the thin disk, we assume a rather short timescale of $\tau_{\rm in}$=0.5 Gyr for the thick disk, and a much longer timescale of $\tau_{\rm in}$=5 Gyr for the thin disk. The metallicity $Z_{\rm in}$ of infalling gas is assumed to be pre-enriched \citep[see][]{Bekki_11b}.  For the thick disk, we set [Fe/H] = -1.3 since the MDF of thick disk stars shows a sharp increase from [Fe/H]$\sim$-1.3 to the peak \citep{Wyse_95}. For the thin disk, the metallicity is determined by an implication from the cosmic evolution of damped Ly$\alpha$ systems \citep{Wolfe_05}, the metallicity of which is around [Fe/H] = -1.5 at the epoch of thin disk formation. Thus, in our models an infalling gas for the thick disk is somewhat more enriched than that for the thin disk, which is reflected in the Ba abundance while an enhanced SN II-like [$\alpha$/Fe] is assumed for both. In the end, the adopted values of $Z_{\rm in}$ are ([Fe/H], [Mg/H], [Ba/H]) = (-1.3, -0.9, -1.1), (-1.5, -1.1, -1.7) for thick and thin disks, respectively. An initial gas of the thick disk formation is set to retain the same abundances as in $Z_{\rm in}$. The star formation rate (SFR) is assumed to be proportional to the gas fraction with a constant coefficient $\nu$ for the duration $\Delta_{\rm SF}$. The higher value of $\nu$=2 Gyr$^{-1}$ for $\Delta_{\rm SF}$= 1.5 Gyr is adopted for the thick disk, compared with $\nu$=0.4 Gyr$^{-1}$ and $\Delta_{\rm SF}$=12 Gyr for the thin disk. For the initial mass function (IMF), we assume a power-law mass spectrum with a slope $x$ of $-1.35$ (a Salpeter's) together with a mass range ($m_l$, $m_u$)=(0.05 \msp, 50 \msp) \citep{Tsujimoto_97}. 

In Figure 2, we show the model results for the evolution of [Mg/Fe] and [Ba/Mg] against [Fe/H] in the thick disk. We see a good agreement with the observations. In particular, a successful reproduction of the [Mg/Fe] feature suggests that a new SN Ia DTD revealed by extragalaxy studies is compatible with the Milky Way case. In the [Ba/Mg]-[Fe/H] diagram, we find an apparent offset between in the observed [Ba/Mg] values between \citet{Bensby_05} and \citet{Venn_04} for both thick and thin disks.  Therefore, the [Ba/Mg] data by \citet{Venn_04} are shifted by +0.15 dex, which is equivalent to a mean difference between  two data sets for -0.5$\leq$[Fe/H]$\leq$-0.2. 

We regard the thick disk as a first disk which is heated up by an ancient minor merger, that is subsequently followed by the gradual formation of a secondary disk, i.e., the thin disk. Such a first thick disk can also be formed through clump merging in an unstable primordial disk \citep{Bournaud_07}. In these scenarios, we assume that star formation within the thin disk occurs after the termination of star formation in the thick disk, and thus the thin disk stars start forming from the thick disk's remaining gas (corresponding to $\sim 10$\% of the original gas) mixed with the gas accreted onto the disk. Accordingly, chemical abundances of the first stars in the thin disk are similar to those of the most metal-rich stars  in the thick disk. As shown in Figure 2,  the thick disk formation leaves the metal-rich gas as an end product of its chemical evolution. As a result, the evolution of [Mg/Fe] starts from [Fe/H]$\sim$0 (Fig.~3). Then, [Fe/H] and [Mg/Fe] decreases and increases, respectively, owing to dilution by metal-poor infalling gas. This reverse evolution comes to an end when the chemical enrichment by star formation exceeds the effect of gas dilution, and subsequently an usual evolutionary path appears. Our claim is that the remaining metal-rich gas after the thick disk formation results in the presence of no metal-poor thin disk stars. In the plot of [Ba/Mg], there are few observed data which coincide with the predicted early evolutionary path until [Fe/H]$\sim$-0.8.  We note that it is not a contradictory result because the predicted fraction of stars riding this path is very small.

\section{Chemical Evolution of Bulge}

We examine the chemical evolution of the Galactic bulge from a viewpoint that it is composed of two different populations. Models are constructed based on two-peaked MDF \citep{Hill_11, Bensby_11}. Our grand picture of their formation is that (i) first, a metal-poor population (MPP) is formed with a relatively short timescale, and (ii) subsequently from its remaining gas plus a gas inflow from an inner disk, a metal-rich population (MRP) starts to form with a longer timescale. For the model of MPP, we assume a pre-enriched initial gas abundance, ([Fe/H], [Mg/H], [Ba/H])=(-1.3,  -0.9, -1.1), so as to reproduce lack of stars for [Fe/H]\ltsim -1.3 in the MDF and high Ba abundances of the most metal-poor stars. Abundances of an infalling gas are assumed to be the same. The enriched gas is considered to be an end result of chemical processing associated with the halo formation. The results calculated with ($\tau_{\rm in}$, $\nu$, $\Delta_{\rm SF}$)=(0.3, 4, 1) together with a Salpeter IMF are shown by blue curves in Figure 4.

\begin{figure}[t]
\vspace{0.2cm}
\begin{center}
\includegraphics[width=7cm,clip=true]{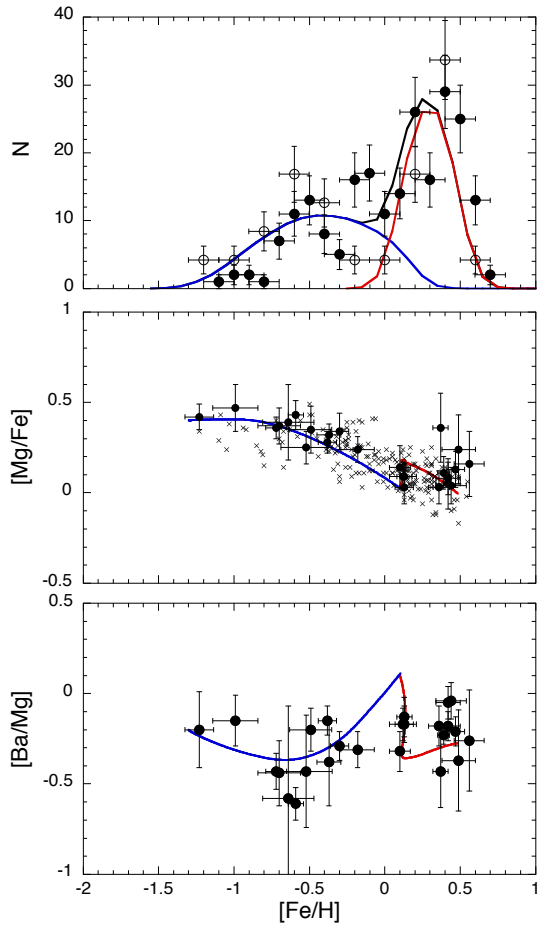}
\end{center}
\vspace{0.3cm}
\caption{Predicted features of chemical evolution for the two-component bulge. The model results for metal-poor and metal-rich components are shown by blue and red curves, respectively. The calculated MDFs are convolved using the Gaussian with a dispersion of 0.1 dex in [Fe/H] considering a measurement error expected in the data. The MDF summing each distribution up with a ratio of 0.5/0.5 is denoted by a black solid curve. The model distribution and the observed one by \citet{Bensby_11} are normalized to coincide with the total number of the sample stars used by \citet{Hill_11}.The observed data are taken from \citet[][filled circles]{Hill_11} and \citet[][open circles]{Bensby_11} for the MDF, \citet[][filled circles]{Bensby_11} and \citet[][small crosses]{Gonzalez_11} for [Mg/Fe], and \citet{Bensby_11} for [Ba/Mg].
}
\end{figure}

MRP is calculated with an initial condition given by a metal-rich gas abundance as an end result of a former MPP's formation under the setting of ($\tau_{\rm in}$, $\nu$, $\Delta_{\rm SF}$)=(1.5, 3, 4). An inflow gas from an inner disk is assumed to be enriched up to ([Fe/H], [Mg/H], [Ba/H])=(-0.3,  -0.3, -0.5). 
This subsolar metallicity is expected in the chemical evolution of the inner disk at an early epoch since at the inner disk an efficient chemical enrichment proceeded but the present-day gas abundance is not so metal-rich such as  [Fe/H]$\sim$ +0.2-0.3 inferred from Cepheids \citep{Andrievsky_04}. This together with the consideration that sufficient time for the release of SN Ia and $s$-process products elapsed results in the above elemental abundances. In addition, in its modeling, we find that a top-heavy IMF is indispensable to make a metal-rich MDF as observed. Otherwise, a peak in the MDF results in being close to [Fe/H]$\sim$0  with an end metallicity unreachable to [Fe/H]$\sim$+0.3. Here we assume $x$=1.05. The results are shown by red curves. In the MDF (top panel), each contribution denoted by colored curves is summed up with an equal ratio according to an observed implication, to construct the predicted MDF (black curve) that can be compared with the observed two-peaked one. We also see that the predicted [Mg/Fe] and [Ba/Mg] evolutions are consistent with the observed features. Note that a gas fraction at the end of star formation for each case of MPP and MRP is 9\%, 8\%, respectively in our calculations. \citet{Cescutti_11} first shows the Ba evolution in the bulge, calculating [Ba/Fe] together with the Mg/Fe evolution. Combination of these two evolutions broadly yields little evolution of [Ba/Mg] ($\sim$-0.4 $-$ -0.5) at least for [Fe/H]\ltsim 0. Its predicted level that is lower than the observed [Ba/Mg] on the whole is likely to be caused by a low Ba yield for the $s$-process in their models.

Our new chemical evolution models have demonstrated that a top-heavy IMF is required to explain the observed MDF and dependences of [Mg/Fe] and [Ba/Mg] on [Fe/H] in the bulge stars. Some previous studies support a top-heavy IMF in the bulge from a different angle. \citet{Dokkum_08} revealed that IMFs in early-type  galaxies at $0.02 \le z \le 0.83$ are significantly flatter than the present-day value of the Galaxy. \citet{Larson_98} discussed a number of items of observational evidence that support top-heavy IMFs in high-redshift spheroidal galaxies. In addition, \citet{Maness_07} claimed a top-heavy IMF in the Galactic center. We find an alternative model to explain the metal-rich part of MDF with the IMF unchanged. The device of this model is an assumed inflow which will chemically evolve to [Fe/H]$\sim$+0.5 during the bulge formation. However, its scenario implicitly presumes a top-heavy IMF in an inner disk outside the bulge. Therefore, we conclude that a top-heavy IMF is unavoidable for the chemical evolution of the inner Galaxy.

In our scenario, metal-rich bulge stars are formed through an inflow from the inner thin disk. On the other hand,    the bulge enriches the disk with large-scale winds to induce the production of metal-rich disk stars \citep{Tsujimoto_10}. In this way, we present a new view on Galactic chemical evolution that an interplay between the bulge and the disk accelerates their chemical enrichment each other.

\section{Discussion}

\subsection{One-component bulge}

The complexity of the stellar population in the Galactic bulge has just entered the stage of debate on its potentiality. The one-component bulge is still a plausible view in terms of the MDF \citep{Fulbright_06, Zoccali_08} or stellar kinematics \citep{Howard_09}. Thus, it is worthwhile to refer to the chemical evolution based on a single-peaked MDF derived from red giants in the Baade window \citep{Fulbright_06}. The reason different types of stars yields different MDFs remains unexplained \citep[][for a detailed discussion]{Bensby_11}. First, we show that the model with an enhanced star formation is not sufficient to reproduce the chemical feature of bulge stars. The dashed curve in the top panel of Figure 5 represents a resultant MDF calculated by the model with a rapid collapse ($\tau_{\rm in}$=0.3 Gyr), a high SFR ($\nu$=2 Gyr$^{-1}$),  and $\Delta_{\rm SF}$=2 Gyr together with a Salpeter IMF. These setting results in a final gas fraction of 10\%. For the metallicity of an initial gas as well as of an infalling gas, we assume a very low-metal content, ([Fe/H], [Mg/H], [Ba/H])=(-3.0,  -2.6, -3.2). The MDF thus obtained is entirely skewed to a low metallicity. In addition, the predicted [Mg/Fe] curve in the middle panel is lower than the observed data in a metal-rich regime. 

\begin{figure}[t]
\vspace{0.2cm}
\begin{center}
\includegraphics[width=7cm,clip=true]{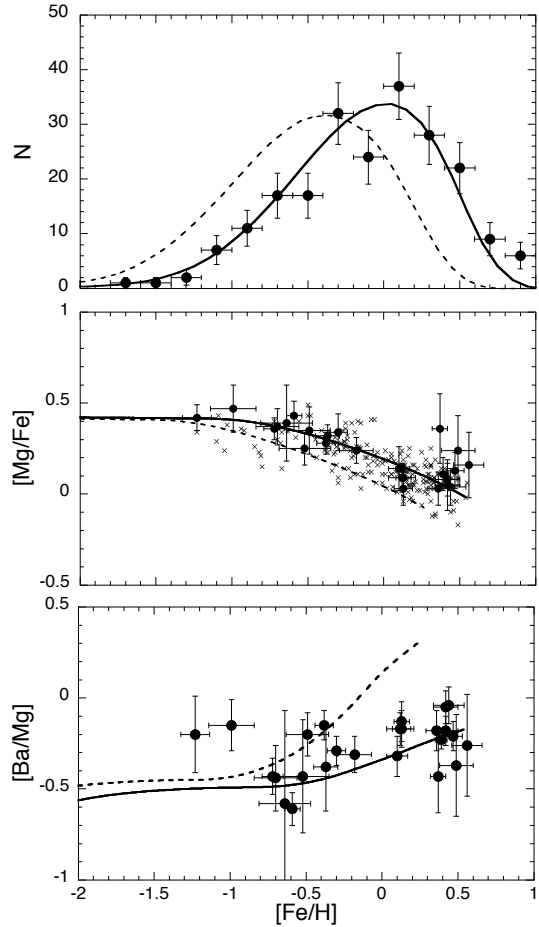}
\end{center}
\vspace{0.3cm}
\caption{Predicted features of chemical evolution including the MDF for the one-component bulge. The results with a flatter IMF ($x$=-1.05) are shown by solid curves while those with a Salpeter IMF ($x$=-1.35) are denoted by dashed curves. The observed data for the MDF is taken from \citet{Fulbright_06}.
}
\end{figure}

On the other hand, the predicted [Ba/Mg] in the bottom panel exhibits a sharper rise from a much lower metallicity than is expected from the observation. These inconsistencies are resolved by the models with a flatter IMF ($x$=1.05), as shown by solid curves, which give a good agreement with the observed MDF as well as the observed correlations of [Mg/Fe] and [Ba/Mg] with [Fe/H]. This flat IMF is fairly consistent with those claimed by several authors, such as $x=1.1$ \citep{Matteucci_90} and $x=0.95$ (Ballero et al.~2007a,b). The necessity of a top-heavy IMF in the bulge can be also concluded from the following argument. An anticipated rapid collapse realizes the condition approximated by a closed-box model. In this approximation, the mean metallicity of stars will become the heavy-element yield when the remaining gas approaches zero \citep{Tinsley_80}. In other words, the location of a peak in the MDF is fundamentally determined by the IMF (not by the SF efficiency).

In the end, our study has demonstrated that a top-heavy IMF is the key factor to efficiently drive chemical enrichment in the bulge, whether one-component or two-components. Recently revealed two-component bulge should be validated by a bigger database of a large sample of bulge stars to confirm the dip around [Fe/H]$\sim$0 in the MDF.

\subsection{Halo vs. short-delayed SNe Ia}

The well-known fact for the Galactic halo is basically no indication of SNe Ia for the elemental abundance of halo stars, which exhibits a plateau of [$\alpha$/Fe] ratio over a whole metallicity range. It turns out that in the CDM Universe halo stars were rapidly formed in the Galactic building blocks, likely in dwarf galaxies, with its termination probably due to huge energy released by numerous SNe II before the major occurrence of of short-delayed SNe Ia. Thus, individual building blocks must be formed with a very short timescale ($\sim 10^8$yr), while an assembly of them finally makes the stellar halo which exhibits an age span of a few Gyr \citep{De Angeli_05}. Observational studies reveal that the chemical abundances of the Galactic stellar halo are significantly different from those of the present-day dwarf galaxies around the Galaxy in the sense that halo stars have higher [alpha/Fe] \citep[e.g.,][]{Shetrone_01, Tolstoy_03, Venn_04}. Recent theoretical calculations based on the $\Lambda$ CDM model show that the majority of the Galactic halo stars are formed from a few relative massive dwarfs in which star formation is rather rapid and thus the stars are  chemically enriched primarily by SNe II to have an enhanced [alpha/Fe] ratio \citep{Robertson_05}. These numerical results should be reanalyzed by introducing the new DTD for SNe Ia.

It would be interesting to check if the $s$-process elements from AGB stars starting to release with a timescale of a few $10^8$yr are imprinted in the abundances of halo stars. In the case where the $s$-process operated in the stellar halo, the [$s$-process/$r$-process] ratio such as [Ba/Eu] or [La/Eu] switches its constant value to an increasing feature from a certain metallicity owing to a gradual $s$-process contribution superimposed on the $r$-process material in the ISM. Observationally, the metallicity indicating this onset among the Galactic halo stars is difficult to detect since it is hidden by a large scatter in the abundance ratios. \citet{Gilroy_88} find the onset of $s$-processing at [Fe/H]$\sim$-2. On the other hand, recent study claims no indication of the $s$-process at least until [Fe/H]$\sim$-1.4 \citep{Roederer_10}. For [Fe/H]$>$-1.4, either the presence or the absence of $s$-processing is very unclear because [La/Eu] continues to make a plateau while [Pb/Eu] shows an upward trend with an increasing [Fe/H]. This issue points to the need for further studies.

If the signature of $s$-processing will be detected among halo stars, how do we connect the new DTD to the chemistry of halo stars? A deficiency of binary stars in the Galactic globular clusters (GCs) compared to the field is reported \citep[e.g.,][]{Pryor_89, Cote_96}. Suppose that this observed fact results from the lack of binaries with short separations in GCs, some of which will be the progenitors of short-delayed SNe Ia in either a DD scenario or a SD scenario. Then, the GC stars are expected to be hardly enriched by SNe Ia.  
These GCs are suggested to be originally formed within galactic building blocks \citep{Searle_78}, and halo stars can be regarded as the stripped stars of the blocks \citep{Bekki_01}. Therefore, we can hypothesize that in individual blocks both GCs and halo stars shared giant molecular clouds, where a lower fraction of close binary stars could be formed due to their high densities, as their birth place. In addition, recent observed result based on chemically unusual red giants implies that a significant fraction of halos stars are originated from disintegrated GCs \citep{Martell_11}. Accordingly, halo stars could avoid the contamination by SNe Ia in their chemical history. The compatibility of the presence of $s$-process among halo stars with the short-delayed DTD can be also understood consistently if $s$-process elements are produced in fast-rotating massive stars \citep{Pignatari_08, Chiappini_11}.

\section{Conclusions}

The new model for the Galactic bulge with two episodes of star formation is presented. We show that it explains the two-peaked MDF recently deduced from red clump stars, that can avoid a contamination of disk stars more reliably than red giants, as well as from microlensed dwarf stars. We stress that this successful MDF reproduction is accompanied by a coincidence of  the observed and predicted evolutions of [Mg/Fe] and [Ba/Mg]. In particular, we first examine the [Ba/Mg] evolution in the bulge, adopting a new Ba yield, which results in the compatibility with an observed level of Ba abundances for both disks and the bulge. Our result will give feedback to the $s$-process nucleosynthesis calculations. We find that an early [Ba/Mg] trend supports a pre-enriched gas for the bulge formation, which is consistent with  lack of metal-poor stars in the observed MDF with two peaks. 

The new DTD for SNe Ia is found to be compatible with the chemical evolution of the Galaxy, leaving room for further discussion on chemical abundances of halo stars. This short-delayed SNe Ia virtually enhance the Fe production with a short timescale together with SNe II. Our study, however, confirms that a top-heavy IMF is still required to explain the chemical feature of the Galactic bulge. Many works thus far have been devoted to the investigation of the effect of various SN Ia DTDs on chemical evolution in the solar vicinity, focusing on the observed [$\alpha$/Fe] break \citep[e.g.,][]{Matteucci_09, Kobayashi_09}. On this matter, we claim that the discussion on the DTD in this context should be assessed by comparing the modeled chemical feature of the thick disk with the corresponding observed one.

\acknowledgements
The authors wish to thank an anonymous referee for his/her valuable comments and excellent review, that has considerably improved the paper. This paper is based upon work supported in part by the National Science Foundation under Grant No.~1066293 and the hospitality of the Aspen Center for Physics. TT is assisted in part by Grant-in-Aid for Scientific Research (21540246) of the Japanese Ministry of Education, Culture, Sports, Science and Technology.

\end{document}